\documentstyle[epsfig]{mn}
\begin{document}

\title[CV evolution: AM Her binaries and the period gap]
{Cataclysmic variable evolution: AM Her binaries and the period gap}
\author[R.F. Webbink \& D.T.~Wickramasinghe]{R. F. Webbink$^{1,2}$
\& D. T. Wickramasinghe$^{1,2}$\\
$^1$ Department of Astronomy, University of Illinois, Urbana-Champaign\\
$^2$ Department of Mathematics, Australian National University, Canberra}

\date{Accepted.  Received}
\pagerange{\pageref{firstpage}--\pageref{lastpage}}
\pubyear{2001}

\maketitle

\begin{abstract}
AM Her variables -- synchronised magnetic cataclysmic variables (CVs) --
exhibit a different period distribution from other CVs across the period gap. 
We show that non-AM Her systems may infiltrate the longer-period end of the
period gap if they are metal-deficient, but that the position and width of the
gap in orbital period is otherwise insensitive to other binary parameters
(excepting the normalisation of the braking rate).  In AM Her binaries,
magnetic braking is reduced as the wind from the secondary star may be trapped
within the magnetosphere of the white dwarf primary.  This reduced braking
fills the period gap from its short-period end as the dipole magnetic moment
of the white dwarf increases.  The consistency of these models with the
observed distribution of CVs, both AM Her and non-AM Her type, provides
compelling evidence supporting magnetic braking as the agent of angular
momentum loss among long-period CVs, and its disruption as the explanation of
the $2^{\rm{h}} - 3^{\rm{h}}$ period gap among nonmagnetic CVs.

\end{abstract}

\section{Introduction}

The disrupted magnetic braking model has enjoyed considerable success in
explaining the $2-3$ hour period gap in the period distribution of the
cataclysmic variables (CVs).  The essential idea is that the CVs with orbital
periods $P_{\rm{orb}} \le 10^{\rm{h}}$ evolve towards shorter periods due to a
combination of angular momentum losses from magnetic braking due to a
magnetised wind originating from the secondary star, and from gravitational
radiation (Verbunt \& Zwaan~\shortcite{vz81}; Rappaport, Verbunt \&
Joss~\shortcite{rvj83}; see Howell, Nelson \& Rappaport~\shortcite{hnr01} for
a recent review).  Magnetic braking dominates at longer periods.  As the donor
secondary decreases in mass, its thermal time-scale increases, until thermal
relaxation can no longer keep pace with the mass transfer rate being driven by
magnetic braking, and it becomes increasingly bloated with respect to thermal
equilibrium.  When, as the orbital period reaches ($\sim\!\!3^{\rm{h}}$), the
secondary's convective envelope penetrates to the stellar center, its
large-scale magnetic field loses its radiative anchor and, being buoyant, is
expelled from the star.  This sudden cessation of magnetic braking allows the
distended secondary to contract more rapidly than its Roche lobe, and mass
transfer ceases until angular momentum losses to gravitational radiation have
reduced the binary separation far enough for the secondary to again fill its
Roche lobe, at an orbital period of $\sim\!\!2^{\rm{h}}$.  It is important to
recognize that, were the secondary to remain in thermal equilibrium throughout
its evolution, it would evolve continuously through the period gap with no
abatement of mass transfer.

A subset of the CVs are known to be magnetic and classified as either AM Hers
or intermediate polars (IPs)~\cite{p94}.  In the AM Hers the magnetic white
dwarf is in synchronous rotation with the orbit so that its spin period
$P_{\rm{spin}}$ is equal to the orbital period $P_{\rm{orb}}$.  In the IPs,
the magnetic white dwarf is asynchronous (typically $P_{\rm{spin}}\sim 0.1
P_{\rm{orb}}$).  Some 65 confirmed AM Hers are known, of which 37 have direct
magnetic field measurements (see Wickramasinghe \& Ferrario~\shortcite{wf00}
for a recent review).  The estimated effective dipolar fields range from
$\sim\!7 - 200$ MG.  The IP class contains fewer confirmed members (26). 
Indirect estimates of their field strengths suggest dipolar fields in the
range $\sim\!5 -30$ MG.  Any theory which purports to explain the period
distribution of CVs must also account for the properties of the magnetic CVs
which, as a group, comprise some $\sim\!\!20 \%$ of all CVs of known orbital
period~\cite{d01}.

The possibility that the magnetic field of the white dwarf may affect the
orbital evolution of the binary was first discussed by Wickramasinghe \&
Wu~\shortcite{ww94} in relation to the AM Her type systems.  This research was
prompted by the early discovery of several AM Herculis systems in the period
gap before the results of the enlarged sample of AM Hers from the ROSAT survey
became available.  Here it was argued that at strong fields, the magnetised
wind from the secondary may become trapped by the combined magnetosphere of
the primary and the secondary and result in a reduction in the magnetic
braking rate.  If the braking rate was effectively curtailed, systems will
evolve towards shorter periods almost entirely by gravitational radiation. 
The proposal was that there will be no disruption in magnetic braking, and
that no period gap will form.

Li, Wu \& Wickramasinghe~\shortcite{lww94a,lww94b,lww95} extended the Mestel
\& Spruit~\shortcite{ms87} wind theory for single stars and developed a theory
for evaluating the braking rate in synchronised AM Herculis type binary
systems.  These models assume that the secondary star has a dynamo generated
field with a dipole moment oriented perpendicular to the orbital plane, and
that the magnetic white dwarf is locked in synchronous rotation by
magnetostatic torques.  The dipole moment of the white dwarf is then oriented
anti-parallel to that of the secondary star.  The calculations showed that for
a fixed orbital period, the braking rates dropped by several orders of
magnitude as the polar field strength increased from $\sim\!\!10$ to
$\sim\!\!100$ MG, and the magnetosphere of the white dwarf engulfed the
companion star and trapped the wind.  Subsequently, Li \&
Wickramasinghe~\shortcite{lw98} allowed for more general orientations of the
dipole moment of the secondary star and showed that the largest reductions in
the braking rate occurred for the case where the dipole moment was nearly
(within $20 \degr$) aligned with the spin axis.

The ROSAT survey~\cite{v99,v00} resulted in a large increase in the number of
known AM Hers~\cite{t98}, and allowed a better statistical analysis of the
period distribution of the magnetic and non-magnetic CVs.  At long orbital
periods, the numbers of AM Hers falls much more rapidly than those of non-AM
Hers, a difference attributable to a competition between accretion torques and
magnetic torques which favors magnetic torques at low donor masses, short
orbital periods, and small orbital separations.  As CVs evolve to shorter
orbital period, then, a growing fraction achieve synchronism and evolve from
non-AM Her to AM Her systems.  Meaningful comparisons of AM Her to non-AM Her
period distributions are therefore limited to short orbital periods, where
statistics are presumably dominated by the influx of systems from longer
orbital periods, as opposed to birth within the period interval of interest. 
Wheatley's~\shortcite{wh95} application of the two-sided Kolmogorov--Smirnov
test to systems with orbital periods less than 3{\fh}2 yielded a $15 \%$
probability that AM Her and non-magnetic distributions were drawn from the
same parent population, a difference he regarded as not significant.  (Hellier
\& Naylor~\shortcite{hn98}, on the other hand, found a similar figure for
systems with orbital period less than $5^{\rm{h}}$, and regarded it as
significant.)  We shall see below that the differences between AM Her and
non-AM Her distributions should be most evident within the period gap itself,
and that a two-sided Kolmogorov--Smirnov test applied only to systems
\emph{within the gap} yields a $8.5 \%$ probability that they belong to the
same parent distribution.  This comparison is still hampered by small numbers,
but we regard the difference as compelling, if not yet conclusive.

In this paper we reports results of calculations of the orbital evolution of
magnetic CVs which allow for a reduction in magnetic braking following the
theory of Li et al.~\shortcite{lww94a}.  We use our results to analyse the
data on the period distribution of the magnetic CVs, and more generally to
re-assess the role played by disrupted magnetic braking in the formation of
the period gap.

\section {Observational constraints}

Magnetic field measurements from Zeeman and cyclotron spectroscopy provide
estimates either of the effective dipolar photospheric field, or of magnetic
fields at accretion spots on the white dwarf surface.  We use as our data base
for AM Hers the list of field measurements given in~\cite{wf00} based on these
techniques updated to include the recent field measurement of V884
Her~\cite{s01}.  The braking calculations, however, require the magnetic
moment which is difficult to estimate from observations. The magnetic moment
while being proportional to the magnetic field is also proportional to the
cube of the stellar radius.  Unfortunately, the latter, which is derived from
the mass, is poorly determined.  Indeed, even in the few eclipsing systems
where the orbital inclination is well constrained, there is no consensus on
the component masses, and these are even less well determined in other
systems.  In our calculations, the magnetic moments have been chosen to be in
the range $10^{32} - 10^{35}$ G~cm$^3$ which is consistent with the observed
effective dipolar field range if allowance is made for a spread in white dwarf
mass.

Our list of CVs is drawn from the database of Downes et al.~\shortcite{d01}
current on 2002 January 25.  We have included the intermediate polars in our
`non-magnetic' sample because to a first approximation, we expect them to
evolve under the normal braking law.  However, it is likely that many IPs
become synchronised as they evolve toward shorter periods, becoming AM Her
systems.  This possibility needs to be considered when comparing observed
distributions.

\begin{figure}
\epsfig{file=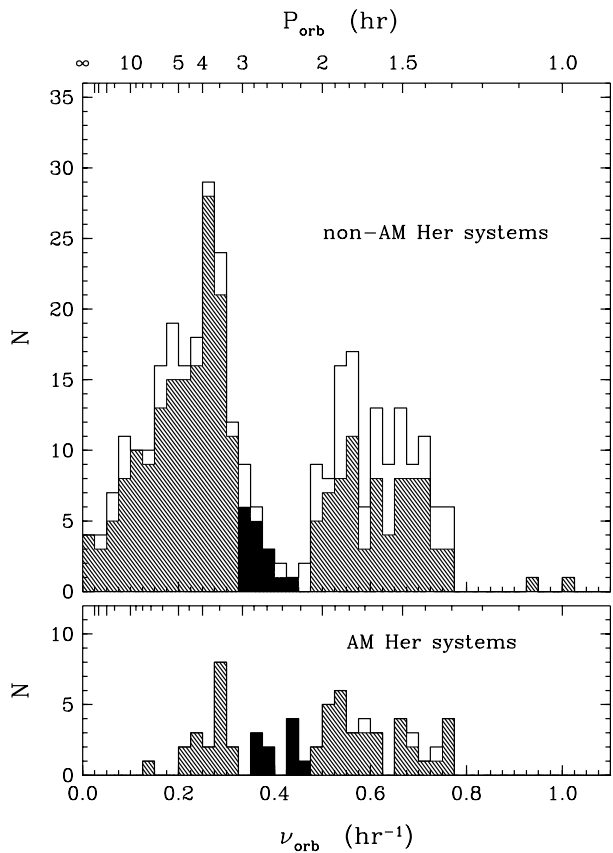,width=0.45\textwidth}
\caption{Orbital frequency distribution of AM Herculis type systems (bottom
panel), and all other types of CVs (top panel).  Filled portions of the
histograms correspond to systems with confirmed orbital periods, with the
black portions highlighting objects in the canonical $2^{\rm{h}} - 3^{\rm{h}}$
period gap.  Open portions of the histograms represent orbital periods of
additional systems considered uncertain/unreliable by Downes et
al.~\shortcite{d01}, or derived therein from superhump periods.}
\end{figure}

We show in Figure 1 the distributions of AM Hers and other types of CVs as a
function of orbital frequency (hr$^{-1}$), binned into frequency intervals
0.025 hr$^{-1}$.  In these units the canonical `period gap' corresponds to the
frequency range $0.325 - 0.475$ hr$^{-1}$.  We choose to display the
distribution in terms of frequency, rather than more conventionally in terms
of period, in order to spread the short-period peak and to make the period gap
more apparent.  In this figure, and throughout the rest of this paper, we will
continue to display results in terms of orbital frequency for ease of
comparison; however, we will cast our discussion in the more familiar terms of
orbital period, and show at the top of each figure the corresponding period
scale.

A glance at Figure 1 shows the following characteristics:

\begin{description}
\item{There is an abrupt drop in the number of non-AM Her systems at periods
above 2.1 hours, which can be considered the lower end of the gap.  At longer
periods, there is a gradual rise in their number distribution, with the
distribution rising rapidly at $\sim\!\!3$ hours, which can be considered the
upper end of the period gap.  This plot clearly shows that the gap is itself
not totally devoid of stars.  Rather there is a tendency for this canonical
period gap to fill in from the higher period end.}

\item{The period gap is decidedly less well-defined among the AM Her stars. 
The average number of systems per unit orbital frequency within the 2.1--3
hour period range is smaller than among systems at shorter period (higher
frequency), but the increase in numbers at longer period (lower frequency) is
much less pronounced than among the non-AM Her systems.  Moreover, among the
magnetic systems, the canonical period gap appears to fill from the lower
period (higher frequency) end, opposite to the behaviour seen for the other
CVs.  The Kolmogorov--Smirnov two-sample statistic, applied only to those
systems of confirmed orbital period falling within the gap in Figure 1, yields
only a $8.5 \%$ probability that AM Her systems and non-AM Her systems are
drawn from the same population.}
\end{description}

\begin{figure}
\epsfig{file=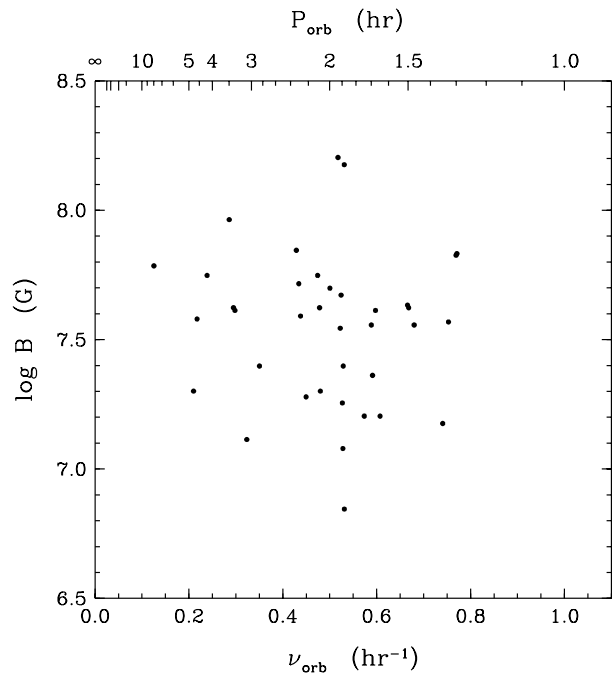,width=0.45\textwidth}
\caption{The logarithm of the magnetic field plotted against orbital frequency
for all AM Hers for which field estimates are available.}
\end{figure}

We show in Figure 2 the distribution of field strength as a function of
orbital frequency for the 37 systems for which field estimates are available. 
The fields correspond to effective dipolar fields for cases where photospheric
Zeeman measurements are available, or where cyclotron fields are available for
both poles.  For other systems, the fields correspond to the value measured at
a single pole.  This plot shows that there is no significant difference
between the mean field of systems within the canonical period gap and outside. 
Furthermore, there is no obvious correlation between field strength and
orbital period for systems within the $2^{\rm{h}} - 3^{\rm{h}}$ period region.

\section{The magnetic braking of synchronised binaries}

\subsection {The parametrisation}
The braking rate of synchronised magnetic binaries depends on the interaction
of the stellar wind with the combined magnetospheres of the components stars. 
It is therefore a function of a number of parameters which fall into two basic
categories: those which specify the magnetic interaction, and those which
specify the properties of the wind.

There are four parameters belonging to the first category which enter directly
into the calculations of the braking rates.  These are the masses $M_1$ and
$M_2$ of the primary and the secondary stars respectively, the radius $R_2$ of
the secondary, and the magnetic moment $\mu_1$ of the primary star.  The
magnetic moment $\mu_2$ of the secondary is not a free parameter but is
calculated on the assumption that prior to the star becoming fully convective,
the mean surface field evolves with angular velocity according to a power law
\[
B_2 = \left( {\omega \over \omega_{\sun}} \right)^{p} B_{\sun}
\]
where $\omega$ is the angular frequency of rotation.  Mestel \&
Spruit~\shortcite{ms87} considered the cases $p=1$ as being most consistent
with the observations of late type stars, and we accordingly adopt $p=1$ in
our models.  However, it should be noted that this law does not allow for
complexities such as dynamo saturation which may play a role in binary
evolution where rapid rotation rates are encountered as the orbit shrinks.

The magnetic moment of the secondary star
\[
\mu_2 = 0.5 B_2 R_2^3
\]
is a strong function of its radius $R_2$.  Among the CV donors we wish to
model, that radius is itself strongly correlated with the stellar mass.  We
therefore parametrise $R_2$ in terms of a pseudo-radius ${\mathcal{R}}_2$
which removes the lowest-order power-law relationship between $R_2$ and
$M_2$~\cite{w95}.  Thus,
\[
R_2 = f {\mathcal{R}}_2 ~ ,
\]
where
\[
{\mathcal{R}}_2 = 0.094 {\rm{R}}_{\sun} (M_2 /0.065 {\rm{M}}_{\sun})^{13/15} ~
.
\]

After the star becomes convective we assume that the intrinsic mean field is
effectively zero.  The secondary star will have an induced dipole moment of 
\[
\mu_2 \simeq \mu_1 (R_2/a)^3
\]
where $a$ is the orbital separation.  The induced dipole moment is typically a
factor $\sim\!10 - 100$ times lower than the intrinsic dipole moment of the
primary star, and there will be no open flux to produce braking even if a wind
persists.  In this regime we can therefore assume that braking is dominated by
angular momentum loss by gravitational radiation.

It can be seen from the above, that apart for the wind parameters to be
discussed in the next section, the braking rate can be parametrised terms of
$M_1,f,\mu_1,M_2$.

\subsection{The calculations of the magnetic braking rate}

In the braking of single stars by a magnetised stellar wind, both thermal and
centrifugal effects can play an important role in different mass and rotation
regimes.  The Mestel \& Spruit~\shortcite{ms87} theory allows a convenient
parametrisation of these effects.  In the context of the CVs, however, due to
their rapid rotation we can assume that centrifugal effects dominate over the
entire observed range~\cite{lww94a,lww94b}.

The formulation of the magnetic braking problem for centrifugally driven winds
has been described in detail by Li et al.~\shortcite{lww94a}.  We therefore
present only the basic equations and our method of solution, and refer the
reader to this paper for further details.
 
The basic assumption underlying our calculations is that the dipole moment of
the secondary star is oriented perpendicular to the orbital plane -- the
likely orientation for a dynamo generated field.  The magnetic moment of the
primary star must then be anti-aligned with that of the secondary in the
minimum energy configuration (stable equilibrium).  The magnetic moment of the
white dwarf rotates through $\pi$ radians as the polarity of the secondary
star switches during its magnetic cycle.  We assume that this adjustment
occurs on a time scale that is shorter than the magnetic cycle time scale, and
that a given system spends the bulk of its evolution with the dipoles nearly
anti-aligned.  In this context we note that in the few well studied AM Hers
there is a preference for the dipole axes to be aligned with the spin
axis~\cite{w95}.  Changes have been observed in the rotational azimuth of the
magnetic axis in three of AM Her systems on a time scale of $\sim\!\!25$ years
(see Wickramasinghe \& Wu 1991, Wu \& Wickramasinghe 1993) but no changes in
magnetic polarity have so far been reported.

In what follows we make the explicit assumption that the braking rates that we
deduce for anti-aligned systems are representative of the mean braking rate
under which the AM Hers evolve.

We consider a coordinate system in which the secondary star is at the origin
$O$, $OZ$ in the direction of the orbital angular momentum, and the primary
star is in the negative $y$ direction at (0,$-a$,0).  The field components in
the plane $OY\!Z$ containing the two dipole moments are
\[
B_y= -3\mu_2 z \left[{y\over r^5}+{k_{\mu}(a+y)\over r_1^5}\right]
\]
and
\[
B_z= \mu_2 \left[ \left( {1 \over r^3}+{3k_{\mu} \over r_1^3} \right) -3z^2
\left( {1 \over r^5}+{3k_{\mu} \over r_1^5} \right) \right]
\]
where 
\[
k_{\mu}={\mu_1\over\mu_2}
\]
where $r_1$ and $r$ are the radial distances of a field point from the primary
and the secondary stars respectively.

The dead zones are assumed to be isothermal with sound speed $c_s^2(d)$ and to
corotate with the secondary star.  The condition of magnetostatic equilibrium
then yields the density within this region.  For a point with polar
coordinates ($r, \theta$) along a field line starting from the surface of the
secondary ($r=R$, $\theta=\theta_0$), the density is~\cite{lww94a}
\begin{eqnarray*}
\rho & = & \rho_0 \ \exp \left[ -l_d \left( 1 - {R \over r} \right) - l_{d1}
\left( 1 - {r_1(R) \over r} \right) \right. \\
& & \left. \mbox{} + {1\over 2} \chi l_d \left( \sin^2 \theta - \sin^2
\theta_0 \right) + \beta \left( {r \over R} \sin \theta - \sin \theta_0
\right) \right]
\end{eqnarray*}
where $\rho_0$ at the density at the base of the dead zone.  Here
\[
\chi l_d = {R^2 \omega^2 \over c_s^2(d)}, ~~~ \beta={M_1 a \over M_1 +
M_2}{R^2 \omega^2 \over c_s^2(d)}
\]
and
\[
\l_d={GM_2\over c_s^2(d) R}, ~~~ \l_{d1}={GM_1 \over c_s^2(d)r_1(R)}
\]
respectively measure the importance of centrifugal and gravitational effects
of the two stars relative to thermal effects on the wind.  We trace the
magnetic field lines by solving the coupled differential equations 
\[
{dy \over ds}={B_y \over B}
\]
and
\[
{dz \over ds}={B_z \over B}
\]
where $B$ is the total field.  The critical field lines which separate the
wind zones from the dead zones are determined by the pressure balance
condition
\[
{B^2 \over 8 \pi}=\rho c_s^2(d) ~ .
\]
For assumed dead zone parameters, which we have evaluated as in Li et
al.~\shortcite{lww94a}, these calculations yield the region $\theta_1 \le
\theta \le \theta_2$ of open field lines in the plane $OY\!Z$ which is then
used to estimate the fractional area $\Phi$ on the surface of open magnetic
flux.

$\Phi$ depends primarily on the magnetic moment of the primary star and of the
mass of the secondary star, and only weakly on $M_1$. The dependence of $\Phi$
on these two parameters is shown in Figure 3.  We note in particular the rapid
drop in open flux for large $\mu_1$ at fixed $M_2$ which plays a crucial role
in determining the characteristics of the orbital evolution to be discussed in
the next section.

As shown by Li et al.~\shortcite{lww94a}, the braking rate takes a
particularly simple form when the wind is centrifugally driven.  In this limit
\begin{eqnarray*}
\dot{J} & = & 4.63 \times 10^{35} {\rm{dyn \ cm}} \left( {\dot{J}_{0} \over
{4.63 \times 10^{35}}} \right) \\
& & \times \left( {5^{\rm{h}} \over P_{\rm{orb}}} \right)^{5/3} \left( {\Phi
\over 0.258} \right)^{5/3} \left( {R_2 \over 0.5} \right)^{10/3}
\end{eqnarray*}
The same theory also provides an expression for $\dot{J}_{0}$ which involves
wind and dead zone parameters.  However, the dependence of these quantities on
spectral type is unknown, and as discussed by Li \&
Wickramasinghe~\shortcite{lw98}, a simple scaling from solar values is
inappropriate.  Following previous investigators, we have accordingly chosen
to normalise the braking rate so as to reproduce the observed period gap in
the limit of zero magnetic field.  This normalisation corresponds to a braking
rate of $4.63 \times 10^{35}$ dyn~cm for a Roche lobe filling secondary star
in a non-magnetic (${\mu}_1 = 0$) CV at an orbital period of
$5^{\rm{h}}$~\cite{rvj83}.

\begin{figure}
\epsfig{file=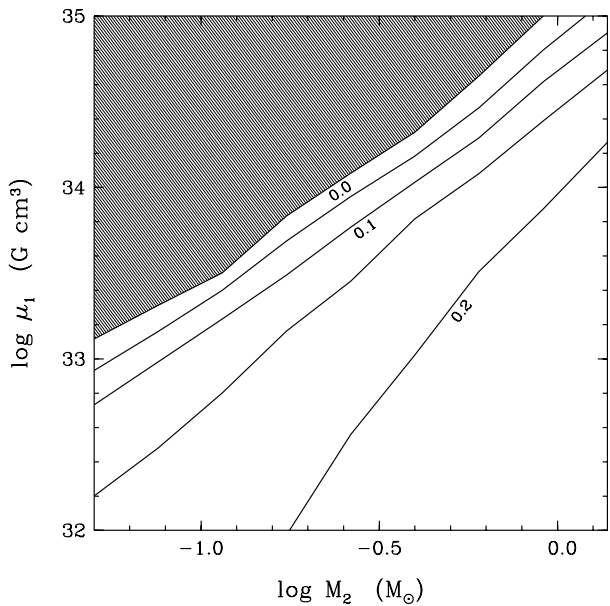,width=0.45\textwidth}
\caption{The fraction $\Phi$ of open field lines as a function of donor star
mass $M_2$ and white dwarf magnetic moment $\mu_1$, for an adopted white dwarf
mass $M_1=0.7$ M$_{\sun}$ and donor radius factor $f=1$.  The shaded region
corresponds to complete closure of the magnetosphere; in this case, magnetic
braking is completely suppressed, and systems evolve only through
gravitational radiation.  $\Phi$ is quite insensitive to $M_1$ and $f$.}
\end{figure}

The mass of the white dwarf and its magnetic moment are essentially free
parameters in our calculations as are the mass of the secondary and the factor
$f$ which determines the radius of the donor star.  We have constructed a grid
of magnetic braking rates as a function of $M_1,f,\mu_1,M_2$, and used these
to calculate orbital evolution for various initial component masses.

\section{evolutionary calculations}

We have explored the evolutionary consequences of attenuated magnetic braking
using the bipolytrope code developed by Rappaport et al.~\shortcite{rvj83}, as
modified by Nelson, Rappaport \& Joss~\shortcite{nrj93}.  This code idealizes
the convective envelope-radiative core structure of near-main sequence stars
of arbitrary degeneracy as $n=3/2$ polytropic envelopes fitted to $n=3$
polytropic cores, and reproduces the most important features of mass-losing
lower main sequence stars with considerable fidelity.  Except as noted below,
we assume solar composition for the donor star throughout, with photospheric
opacities adopted from Alexander~\shortcite{a75,a80} after Rappaport, Joss \&
Webbink~\shortcite{rjw82}.

For purposes of computing mass transfer rates, we fit isothermal atmospheres
to the bipolytropes, and write the mass transfer rate explicitly in terms of
the difference between the fitting point radius for this atmosphere and the
Roche lobe radius of the donor star.  However, in the calculations presented
here, that difference never exceeds $2 \times 10^{-3}$ of the stellar radius,
so we could have achieved the same results for all practical purposes by
limiting the stellar radius to the Roche lobe radius as a surface boundary
condition.

\begin{figure}
\epsfig{file=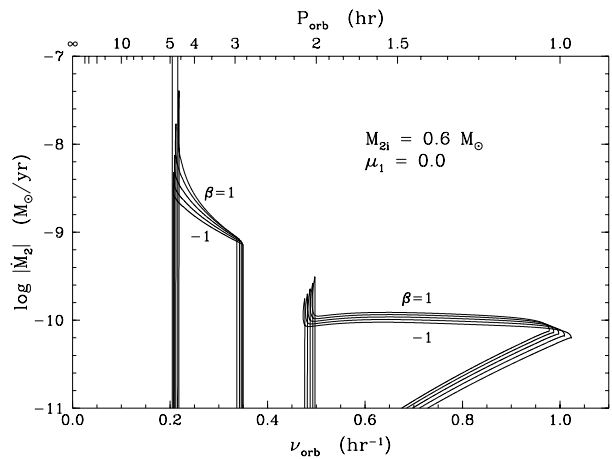,width=0.45\textwidth}
\caption{Mass transfer rates as functions of orbital frequency for different
degrees of systemic mass loss.  The parameter $\beta (=1.0, 0.5, 0.0, -0.5,
-1.1)$ is the fraction of mass lost by the donor which is retained by the
white dwarf.  Where $\beta=1$, mass transfer is completely conservative, while
for $\beta=-1$ the white dwarf is presumed to expel twice the mass it accretes
over a nova outburst cycle.  Our assumed value ($\beta=0$) is emphasized.  The
initial donor mass is fixed at $M_{2i}=0.6$ M$_{\sun}$, and the initial white
dwarf mass adjusted to yield the same total mass ($M_1 + M_2 = 1.268$
M$_{\sun}$) at the upper edge of the period gap in all of the sequences
shown.}
\end{figure}

We assume that the white dwarf mass remains constant throughout the evolution
of our model binaries, with all mass lost by the hydrogen-rich donor lost from
the system in nova explosions.  The ejected mass is assumed to carry off a
specific angular momentum equal to the orbital angular momentum per unit mass
of the white dwarf.  Observationally, nova outbursts are found in non-AM Her
systems and AM Her systems alike (\emph{viz.} V1500 Cyg).  The heavy-element
enhancements typical of the ejecta of well-studied novae~\cite{g98} are
unmistakable evidence that the underlying white dwarfs do not generally grow
in mass during the course of evolution, but may in fact suffer net erosion
through nova outbursts.  Provided that this erosion is not so severe as to
drive the binary toward dynamical instability (roughly, that $dM_1/dM_2<3/2$),
that the binary in question has not arrived at the onset its cataclysmic
variable phase just above the period gap, and that the angular momentum
content of nova ejecta has not been seriously underestimated, then the width
and position of the period gap are insensitive to our assumption of constant
white dwarf mass.  Figure 4 illustrates this point, by comparing a family of
evolutionary tracks for nonmagnetic CVs, differing in the extent of systemic
mass loss, but converging to identical total masses at the upper edge of the
period gap.

\begin{figure}
\epsfig{file=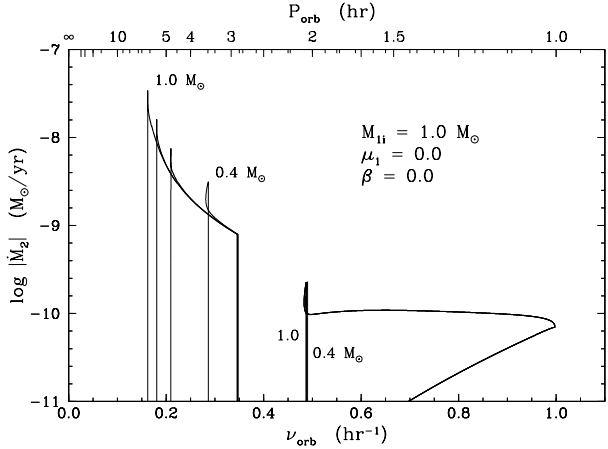,width=0.45\textwidth}
\caption{Mass transfer rates as functions of orbital frequency for systems
differing only in initial donor star mass ($M_{2i}=1.0, 0.8, 0.6, 0.4$
M$_{\sun}$).  The emphasized curve corresponds to the same initial conditions
($M_{1i}=1.0$ M$_{\sun}$, $M_{2i}=0.6$ M$_{\sun}$, $\mu_1$=0, $\beta=0$, and
solar metallicity, [Fe/H]=0, throughout Figures 4-8.}
\end{figure}

Figure 5 underscores the point implicit in Figure 4: for a given magnetic
braking rate, the position and width of the period gap depends principally on
the component masses of the binary as it ceases magnetic braking, and not on
its evolutionary history in reaching that point.  Since the disrupted magnetic
braking model for the period gap associates the cessation of magnetic braking
with the donor star becoming fully convective, the principal factors governing
the position and width of the period gap (again, for a given magnetic braking
rate) are (i) the mass of the white dwarf (or, equivalently, the mass ratio),
and (ii) the chemical composition of the donor star.

\begin{figure}
\epsfig{file=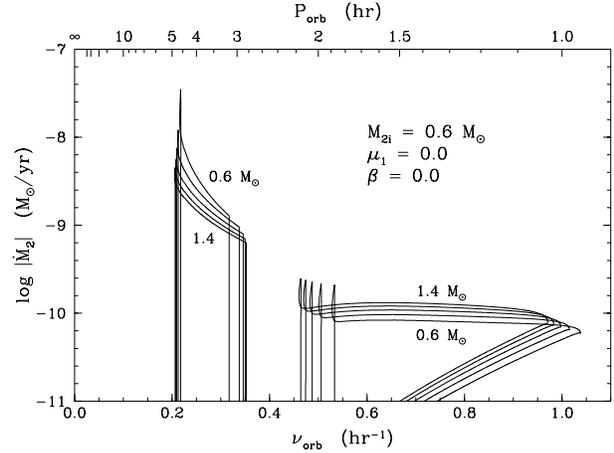,width=0.45\textwidth}
\caption{Mass transfer rates as functions of orbital frequency for systems
differing only in initial white dwarf mass ($M_{1i}=1.4, 1.2, 1.0, 0.8, 0.6,
0.4$ M$_{\sun}$).  The emphasized curve is common to Figures 4-8.}
\end{figure}

The mass of the white dwarf influences the period gap through its role in
orbital dynamics.  Smaller white dwarf masses demand larger mass transfer
rates to moderate the rate of Roche lobe contraction induced by angular
momentum losses to magnetic braking.  These higher mass transfer rates in turn
drive the donor star further from thermal equilibrium, expanding the period
gap, as seen in Figure 6.  However, the effect is only strong for binary mass
ratios near the limit of dynamical stability for fully convective donors
($(M_1/M_2)_{\rm crit}=3/2$), and cannot by itself account for CVs found in
the gap.

\begin{figure}
\epsfig{file=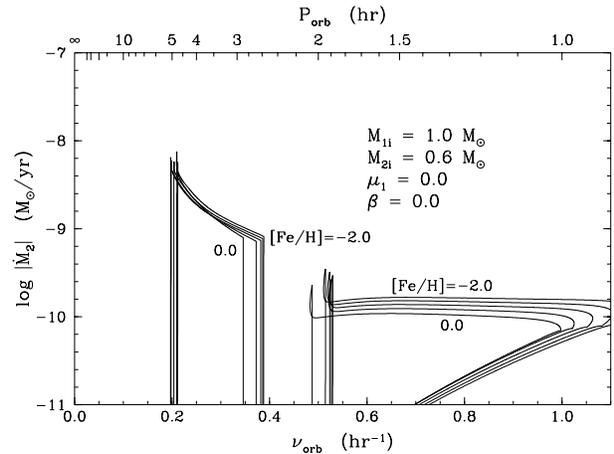,width=0.45\textwidth}
\caption{Mass transfer rates as functions of orbital frequency for systems
differing only in donor star metallicity ([Fe/H]$ = 0.0, --0.5, --1.0, --1.5,
--2.0$).  The emphasized curve is common to Figures 4-8.}
\end{figure}

Figure 7 illustrates the effect of varying metallicity of the donor star on
the position and width of the period gap.  Metal-deficient donors are
systematically bluer and more luminous than their solar-composition
counterparts of the same mass.  Their higher luminosities endow them with
shorter thermal time scales, and they therefore adhere more closely to the
thermal equilibrium main sequence before turning fully convective than do
solar-composition stars.  As seen in Figure 6, then, metal-deficient stars
encroach into the period gap from the long-period (low-frequency) end.  We
tentatively suggest, therefore, that the small number of non-AM Her-type CVs
spilling into the period gap at long period may be relatively metal-poor. 
Metal-deficient main sequence stars are also physically smaller at low masses
than solar composition stars, and so the short-period (high-frequency)
boundary of the period gap migrates toward shorter periods (higher
frequencies).

The preceding considerations demonstrate that, within the context of the
disrupted magnetic braking model, the period gap is a robust entity.  One can
account for small numbers of non-AM Her CVs in the gap by appealing to
newly-formed systems in which the donor stars are already fully convective
main sequence stars with masses in the 0.2--0.3 M$_{\sun}$ range, or by
appealing to infiltration by metal-deficient donors at the long-period end of
the gap.  However, in light of the robustness of the period gap, the
difference in orbital period distributions between AM Her systems and non-AM
Her systems seen in Figure 1 strongly suggests that the AM Her systems deviate
from the disrupted braking law which seems to apply to all other CVs.

As described in Section 3, in synchronous rotators (AM Her binaries), the
dipole field of the white dwarf can close field lines from the donor star,
suppressing the stellar wind on otherwise open field lines, and thereby
suppress angular momentum losses.  We have incorporated the attenuated
magnetic braking model described there into the evolutionary model described
above, under the assumption that the white dwarf magnetic dipole moment, like
its mass, remains constant in time.  Our model sequences assume that the white
dwarf is locked in synchronism with the binary orbit from the very onset of
mass transfer, i.e., that the system is of AM Her type throughout its
evolution.

\begin{figure}
\epsfig{file=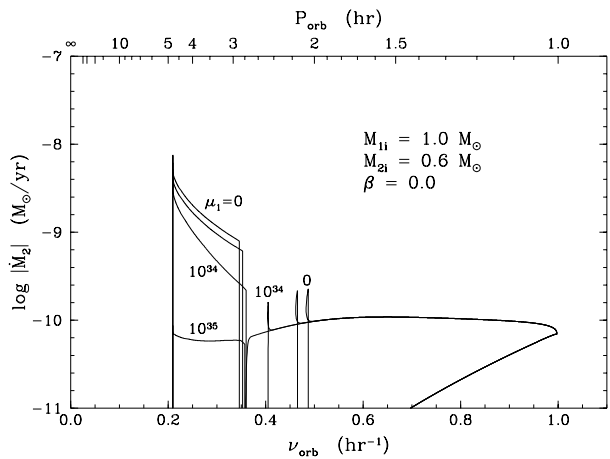,width=0.45\textwidth}
\caption{Mass transfer rates as functions of orbital frequency for systems
differing only in white dwarf magnetic moment ($\mu_1 = 0,10^{33}, 10^{34},
10^{35}$ G cm$^3$).  The emphasized curve is common to Figures 4-8.}
\end{figure}

We show in Figure 8 the effect of increasing magnetic moment, ${\mu}_1$, of
the white dwarf primary on the evolution of a typical AM Her system with
initial component masses $M_1=1.0$ M$_{\sun}$ for the white dwarf and
$M_2=0.6$ M$_{\sun}$ for the donor secondary.  As the braking torque is
suppressed with increasing ${\mu}_1$, mass transfer rates above the period gap
subside, and the donor star is no longer driven so far from thermal
equilibrium before it becomes fully convective at the top (lower frequency
edge) of the period gap.  As the donor star is no longer so severely inflated
in comparison with its thermal-equilibrium main sequence counterpart, so its
contraction following cessation of the magnetic braking driving mass transfer
is also less severe.  Mass transfer resumes at longer orbital period (lower
orbital frequency) than in the disrupted magnetic braking model for non-AM
Her-type CVs because the donor is no longer so severely undermassive for its
orbital period, and the period gap fills from its shorter period (higher
frequency) end.  As seen in Figure 5, this filling of the period gap is
progressively greater as ${\mu}_1$ increases, and magnetic braking prior to
the secondary's loss of its dipole field is suppressed.  For sufficiently high
values of ${\mu}_1$ (see ${\mu}_1 = 10^{35}$ G cm$^3$ in Figure 5), magnetic
braking becomes inconsequential, and the system evolves more or less
continuously through the period gap (the momentary dip in mass transfer rate
at $P \approx 3^{\rm{h}}$ owing to an inflection point in the thermal
equilibrium mass-radius relation at the point stars become fully convective).

\begin{figure}
\epsfig{file=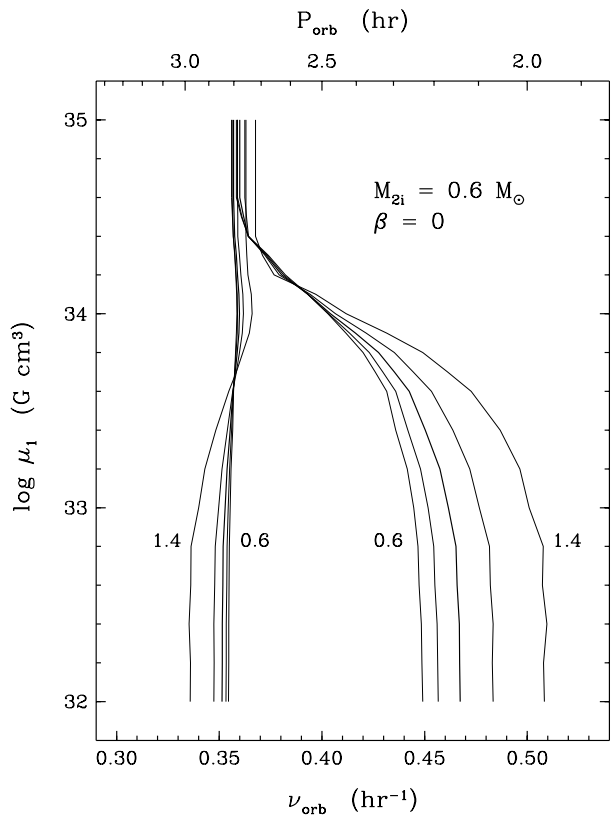,width=0.45\textwidth}
\caption{Upper and lower limits to the period gap, as functions of white dwarf
mass $M_1$ (as labeled, in units of M$_{\sun}$) and white dwarf magnetic
moment $\mu_1$.  The gap boundaries depend on $M_{\rm 2i}$ only for initial
masses $M_{2i}<0.4$ M$_{\sun}$.  $\beta=0$ is assumed.}
\end{figure}

Figure 9 illustrates the dependence of the upper and lower edges of the period
gap on white dwarf mass $M_1$ and magnetic moment ${\mu}_1$.  We define the
upper (lower-frequency) edge of the period gap as the point at which the donor
star becomes fully convective.  In those cases where magnetic braking remains
strong enough to drive the donor out of thermal equilibrium, this point indeed
marks a local minimum in the orbital period, which increases momentarily (if
only by a very small increment) as the mass transfer rate subsides.  We take
the lower (higher-frequency) edge of the period gap as the period maximum
(frequency minimum) during the thermal time scale mass transfer episode
visible as a slim vertical loop in Figure 5.  For sufficiently large white
dwarf magnetic moments, where magnetic braking is nearly completely
suppressed, this loop disappears, and we then define the lower edge of the
period gap as the local maximum in $\ddot{M}_2$.

It is evident from Figure 9 that, for white dwarf moments ${\mu}_1 \la
10^{33}$ G cm$^3$, the gap edges asymptotically approach those of the non-AM
Her systems.  At the other extreme, for white dwarf moments ${\mu}_1 \ga 4
\times 10^{34}$ G cm$^3$, the gap is reduced to a momentary hiccup as the
donor becomes fully convective.  For intermediate white dwarf moments, the
period gap fills from its short-period (high-frequency) end, as indeed
observations suggest (see Figure 1).  A clear prediction of our model is that,
as more AM Her systems are discovered within the period gap, a clear trend of
increasing magnetic moment with increasing orbital period (decreasing orbital
frequency) should emerge.

\begin{figure}
\epsfig{file=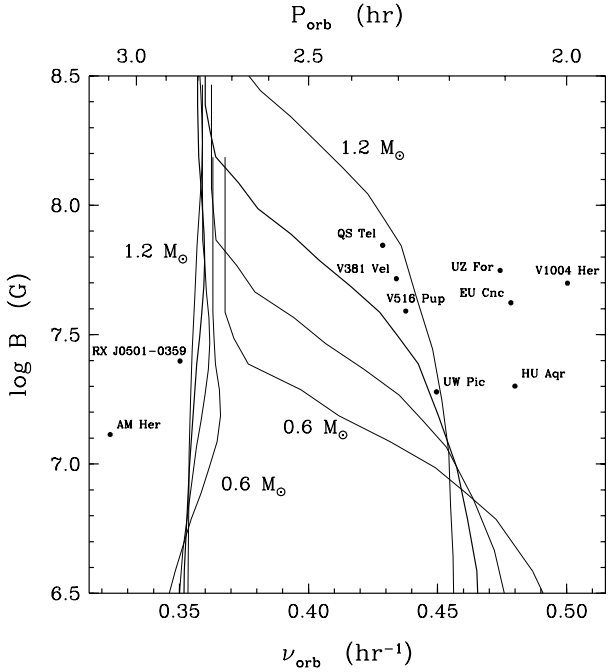,width=0.45\textwidth}
\caption{Observed field strengths of AM Her stars as functions of orbital
frequency, in the region near the period gap.  The continuous curves
correspond to model calculations (cf. Fig. 9) with differing white dwarf mass
$M_1$ ($ = 1,2, 1.0, 0.8, 0.6$ M$_{\sun}$), and enclose the period gap
corresponding to the respective value of $M_1$.}
\end{figure}

To facilitate comparison with observed AM Her stars in the period gap, we have
used the Nauenberg~\shortcite{n72} mass-radius relation for cold white dwarfs
to convert the dipole moments of Figure 9 to polar fields.  The region
immediately surrounding the period gap is shown in Figure 10.  Unfortunately,
empirical masses of white dwarfs in AM Her binaries are generally very
poorly-constrained.  However, to the extent that our model results can be
taken at face value, the positions of observed systems in Figure 10 can be
used to set upper limits to their white dwarf masses, since for given dipole
moment, the polar field strength decreases rapidly with increasing mass.  In
this figure, AM Her binaries must fall \emph{outside} the gap corresponding to
their white dwarf masses, thus implying that QS Tel, V381 Vel, V516 Pup, and
UW Pic must contain white dwarfs of mass $M_1 \la 1.1$ M$_{\sun}$.  The reader
is cautioned, however, that this constraint is quite sensitive to the precise
location of the short-period (high-frequency) edge of the canonical period
gap.

\section{Discussion and conclusions}

\begin{figure}
\epsfig{file=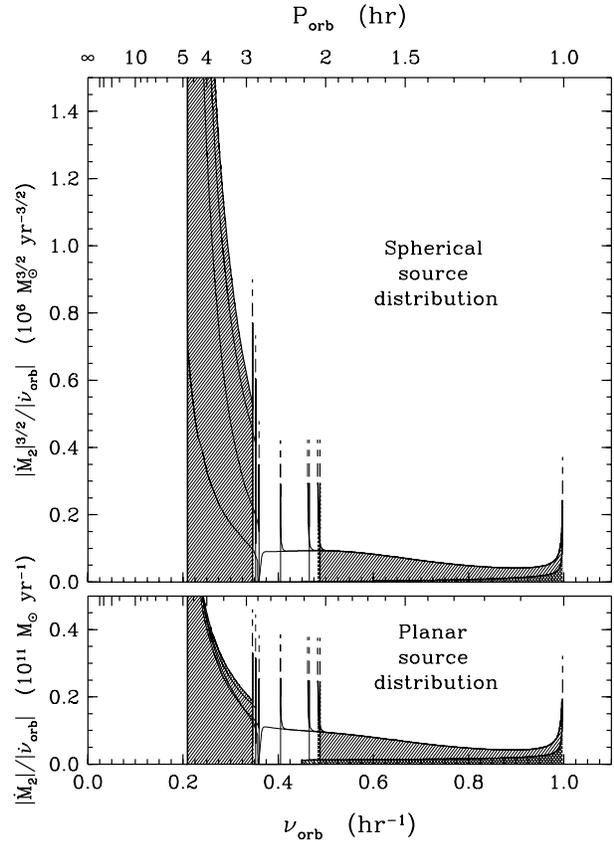,width=0.45\textwidth}
\caption{Relative discovery probabilities for AM Her stars as a function of
orbital frequency.  Only results for our fiducial initial model ($M_1 = 1.0$
M$_{\sun}$, $M_{2i} = 0.6$ M$_{\sun}$, $\beta = 0$, solar metallicity) are
shown, for different values of white dwarf magnetic moment $\mu_1$, as in
Figure 8.  A constant formation rate over the age of the galactic disc
(10$^{10}$ yr) is assumed.  The nonmagnetic case is hatched, with the hatching
sloping upward in the direction of evolution.  The vertical dashed segments
mark singularities which arise in the distributions shown where
$\dot{\nu}_{\rm{orb}}$ changes sign.  Such singularities occur in pairs at the
upper and lower edges of the period gap, but only those at the lower
(higher-frequency) edge of the gap are resolvable here.}
\end{figure}

The AM Her stars, unlike cataclysmic variables generally, have mostly been
discovered through all-sky surveys, where they appear as strong X-ray
emitters.  We can therefore approximate the selection effects governing their
discovery by treating them as an X-ray-brightness limited sample, with X-ray
luminosities proportional to the mass transfer rate $|\dot{M}_2|$.  To the
extent that X-ray extinction is negligible, the distance to which systems are
sampled is then proportional to $|\dot{M}_2|^{1/2}$.  If the threshold
distance is small compared with the scale height of AM Her systems with
respect to the galactic plane, then the apparent distribution should be
roughly isotropic, and the number of systems sampled varies as
$|\dot{M}_2|^{3/2}$; should the distance be large compared with the scale
height (but still small compared with the radial scale length of the galactic
disk) the number of systems varies as $|\dot{M}_2|$ itself.  The number of
systems in any orbital frequency interval is then simply proportional to the
lifetime of the system in that interval, i.e., to $\dot{\nu}_{\rm{orb}}^{-1}$. 
As a point of reference, the mean value of $\sin |b|$ of the AM Her systems in
Figure 1 is 0.51, indistinguishable from that of a spherical source
distribution ($\langle \sin |b| \rangle = 0.50$).

Figure 11 illustrates the relative discovery probabilities arising from this
simple model of selection effects, in the two limiting cases regarding their
spatial distribution, for our fiducial evolutionary sequence.  In reality, we
expect CVs to form with a range of component masses, and this will affect the
overall form of the true distribution.  In particular, systems with a given
initial secondary mass, $M_{\rm 2i}$, contribute to the distribution only at
orbital frequencies $\nu_{\rm{orb}}$ above the frequency at which the CV first
reaches interaction (cf. Figure 5).  Likewise, magnetic CVs will contribute to
the AM Her distribution only at orbital frequencies above those at which they
first reach synchronism.  Nevertheless, as seen above, the properties of
individual systems at the upper and lower edges of the period gap, and the
positions of those edges, are insensitive to their initial conditions.

One feature immediately evident from Figure 11 is that the number density per
unit orbital frequency of CVs at the upper (low-frequency) edge of the period
gap is decidedly greater than at its lower (high-frequency) edge.  The ratio
of these two number densities, as defined at the very edges of the period gap,
amounts to a factor of 5.4 for non-AM Her systems with a spherical source
distribution (1.9 in a planar distribution).  Among the AM Her systems of
increasing magnetic moment ${\mu}_1$, the amount by which this ratio exceeds
unity is roughly proportional to the width of the period gap appropriate to
that value of ${\mu}_1$.

Reference to Figure 1 shows that the non-AM Her systems have a much higher
number density of systems at the upper edge of the period gap than at its
lower edge, in at least qualitative agreement with Figure 11.  However, among
AM Her systems themselves, this ratio of number densities is plainly less than
unity.  The period gap is for them only partially filled, and the marked
increase in number density beyond the lower (high-frequency) edge of the gap
should be matched by an equally marked increase at the upper (low-frequency)
edge.  The clear implication is that more AM Her-type systems emerge from the
period gap than enter it.  Thus, a significant fraction of AM Her systems
below the period gap must have entered the gap in a non-AM Her (asynchronous)
state, i.e., as intermediate polars.  As asynchronous systems, they suffered
full magnetic braking above the gap, and were driven out of thermal
equilibrium to the same extent as other non-AM Her systems.  They therefore
remain detached until resuming mass transfer at the lower edge of the gap as
defined by other non-AM Her systems, in transit becoming synchronously locked,
but failing to appear either as CVs or AM Her systems within the gap.  Systems
which achieve synchronism while crossing the period gap thus contribute to the
AM Her population only at periods below the gap.

Comparison of Figure 11 with Figure 1 also shows that the theoretical period
minimum of AM Her stars lies at significantly shorter period (higher
frequency) than observed.  This is a problem pervading CV evolution generally,
since both AM Her systems and non-AM Her systems are driven through this
period minimum by gravitational radiation.  Kolb \& Baraffe~\shortcite{kb99}
discuss a number of possible explanations for this discrepancy.  Among these
possibilities, they note that a four-fold increase in orbital angular momentum
loss rate (above that given by gravitational radiation alone) would raise the
CV minimum period to that observed (if we discount the anomalous cases of V485
Cen at $P_{\rm{orb}}$ = 0{\fh}984 and 1RXS J2329+0628 at $P_{\rm orb}$ =
1{\fh}070).  However, we find that at these ultashort orbital periods, the
white dwarf magnetic field should suppress a magnetic stellar wind entirely
among AM Her systems.  The fact that both AM Her binaries and non-AM Her
binaries are subject to the same minimum period (Figure 1) then suggests that
stellar winds are not responsible for this discrepancy.

We summarise our principal conclusions as follows:

(a) The width of the period gap, and the position of its lower boundary, is a
function of the magnetic moment of the primary star.  For sufficiently high
magnetic moments ($\mu_1 \ga 4 \times 10^{34}$ G cm$^3$), the gap disappears,
and mass transfer is driven primarily by gravitational radiation.

(b) Observations of AM Her-type systems show that the period gap is less
pronounced than that of CVs belonging to all other classes.  In particular,
the AM Hers appear to fill the gap from the short-period ($\sim\!\!2^{\rm h}$)
end, behaviour which supports the reduced braking theory outlined in section
3.  In contrast, non-AM Her systems encroach on the period gap from its
long-period edge, behaviour which we suggest may reflect lower metallicities
of these particular systems.

(c) The differences between the distributions of AM Her binaries and non-AM
Her binaries within the period gap in turn provide strong prima facie evidence
that magnetic stellar winds are indeed the dominant angular momentum loss
mechanism among non-AM Her binaries.

(d) The abundance of AM Hers immediately below the period gap, in comparison
with their numbers immediately above the period, requires that a significant
proportion of the AM Hers found below $\sim\!\!2^{\rm{h}}$ entered the period
gap as IPs, and become synchronised only as mass transfer ceased within the
gap.  The smaller accretion torques, and increased magnetic interaction (due
to the reduction in orbital separation) allowed these systems to remain
synchronised as they emerge from the gap.

\section{Acknowledgements}

RFW thanks the Astrophysical Theory Centre, Department of Mathematics,
Australian National University, where this work was begun, for its
hospitality, and Lilia Ferrario for her generous assistance.  DTW in turn
acknowledges the hospitality of the Department of Astronomy, University of
Illinois, where this work was completed.  Both authors thank an anonymous
referee for constructive remarks.  Studies of interacting binaries are
supported at the University of Illinois by National Science Foundation grant
AST 9618462.

\end{document}